\documentclass[preprint,superscriptaddress, showpacs,preprintnumbers,amsmath,amssymb]{revtex4}

\allowdisplaybreaks
\allowdisplaybreaks[4]
\usepackage{CJK}
\usepackage{graphicx}% Include figure files
\usepackage{dcolumn}% Align table columns on decimal point
\usepackage{bm}% bold math
\usepackage[dvipdfm,
            pdfstartview=FitH,
            CJKbookmarks=true,
            bookmarksnumbered=true,
            bookmarksopen=true,
            colorlinks=true,
            pdfborder=001,
            linkcolor=blue,
            anchorcolor=blue,
            citecolor=blue
            ]{hyperref}
\begin{document}

\begin{CJK}{GBK}{song}

\title{Wobbling motion in $^{135}$Pr within a collective Hamiltonian}

\author{Q. B. Chen}
\affiliation{State Key Laboratory of Nuclear Physics and Technology,
             School of Physics, Peking University, Beijing 100871, China}%

\author{S. Q. Zhang}\email{sqzhang@pku.edu.cn}
\affiliation{State Key Laboratory of Nuclear Physics and Technology,
             School of Physics, Peking University, Beijing 100871, China}%

\author{J. Meng}\email{mengj@pku.edu.cn}
\affiliation{State Key Laboratory of Nuclear Physics and Technology,
             School of Physics, Peking University, Beijing 100871, China}%
\affiliation{School of Physics and Nuclear Energy Engineering,
             Beihang University, Beijing 100191, China}%
\affiliation{Department of Physics, University of Stellenbosch,
             Stellenbosch, South Africa}%

\date{\today}

\begin{abstract}

The recently reported wobbling bands in $^{135}$Pr are investigated
by the collective Hamiltonian, in which the collective parameters,
including the collective potential and the mass parameter, are
respectively determined from the tilted axis cranking (TAC) model
and the harmonic frozen alignment (HFA) formula. It is shown that
the experimental energy spectra of both yrast and wobbling bands are
well reproduced by the collective Hamiltonian. It is confirmed that
the wobbling mode in $^{135}$Pr changes from transverse to
longitudinal with the rotational frequency. The mechanism of this
transition is revealed by analyzing the effective moments of inertia
of the three principal axes, and the corresponding variation trend
of the wobbling frequency is determined by the softness and shapes
of the collective potential.

\end{abstract}

  \maketitle

%%%%%%%%%%%%%%%%%%%%%%%%%%%%%%%%%%%%%%%%%%%%%%%%%%%%%%%%%%
%                    begin  introduction
%%%%%%%%%%%%%%%%%%%%%%%%%%%%%%%%%%%%%%%%%%%%%%%%%%%%%%%%%%

\section{Introduction}\label{sec1}

The triaxial shape has been a long-standing subject in nuclear
physics. The appearance of the wobbling bands~\cite{Bohr1975,
Odegard2001PRL} and the chiral doublet
bands~\cite{Frauendorf1997NPA, Starosta2001PRL} has provided
unambiguous experimental evidence of triaxiality. The wobbling mode
was first proposed by Bohr and Mottelson in the
1970s~\cite{Bohr1975}. It exists in a triaxial nucleus when the
total spin of the nucleus does not align along any of the principal
axes, but precesses and wobbles around one of the axes, in analogy
to an asymmetric deformed top~\cite{Landau1960book}.

The wobbling bands were first observed in
$^{163}$Lu~\cite{Odegard2001PRL, Jensen2002PRL}. Since then, seven
more wobbling nuclei have been reported, including
$^{161}$Lu~\cite{Bringel2005EPJA},
$^{165}$Lu~\cite{Schonwasser2003PLB}, $^{167}$Lu~\cite{Amro2003PLB},
and $^{167}$Ta~\cite{Hartley2009PRC} in $A\sim 160$,
$^{135}$Pr~\cite{Matta2015PRL} in $A\sim 130$, and even-even
$^{112}$Ru~\cite{S.J.Zhu2009IJMPE} and
$^{114}$Pd~\cite{Y.X.Luo2013proceeding} in the $A\sim 110$ mass
regions. Among the odd-$A$ wobblers, $^{135}$Pr is the only one out
of the $A\sim 160$ mass region, which is built on a proton
$h_{11/2}$ configuration with a moderate deformation ($\beta\sim
0.17$), while the others in $A\sim 160$ involve a proton $i_{13/2}$
configuration with significantly large deformation ($\beta\sim
0.40$).

The excitation energy of a wobbling motion is characterized by
wobbling frequency. In the originally predicted wobbler for a pure
triaxial rotor (simple wobbler)~\cite{Bohr1975}, the wobbling
frequency increases with spin. However, decreasing wobbling
frequencies with spin were observed in the Lu and Ta isotopes as
shown in Ref.~\cite{Hartley2011PRC}. To clarify this contradiction,
Frauendorf and D\"{o}nau~\cite{Frauendorf2014PRC} distinguished two
types of wobbling motions, \emph{longitudinal} and \emph{transverse}
wobblers, for a triaxial rotor coupled with a high-$j$
quasiparticle. For the longitudinal wobbler, the quasiparticle
angular momentum and the principal axis with the largest moment of
inertia (MOI) are parallel; for the transverse one, they are
perpendicular. They demonstrated that the wobbling frequency of a
longitudinal wobbler increases with spin, while that of a transverse
one decreases with spin~\cite{Frauendorf2014PRC}. Therefore, the
wobbling bands in the Lu and Ta isotopes are interpreted as
transverse wobbling bands.

Theoretically, the triaxial particle rotor model
(PRM)~\cite{Bohr1975, Hamamoto2002PRC, Hamamoto2003PRC,
Tanabe2006PRC, Tanabe2008PRC, Tanabe2010PRC, Tanabe2014PTEP,
Frauendorf2014PRC, W.X.Shi2015CPC} and the cranking model plus
random phase approximation (RPA)~\cite{Marshalek1979NPA,
Shimizu1995NPA, Matsuzaki2002PRC, Matsuzaki2004EPJA,
Matsuzaki2004PRC, Matsuzaki2004PRCa, Shimizu2005PRC, Shimizu2008PRC,
Shoji2009PTP, Frauendorf2015PRC} have been widely used to describe
the wobbling motion. Recently, based on the cranking mean field and
treating the nuclear orientation as collective degree of freedom, a
collective Hamiltonian was constructed and applied for the
chiral~\cite{Q.B.Chen2013PRC} and wobbling
modes~\cite{Q.B.Chen2014PRC}. Usually, the orientation of a nucleus
in the rotating mean field is described by the polar angle $\theta$
and azimuth angle $\varphi$ in spherical coordinates. In the
collective Hamiltonian for wobbling modes, the azimuth angle
$\varphi$ is taken as the collective coordinate since the motion
along the $\varphi$ direction is much easier than in the $\theta$
direction~\cite{Q.B.Chen2014PRC}. The quantum fluctuations along
$\varphi$ are taken into account to go beyond the mean-field
approximation. Using this model, the simple, longitudinal, and
transverse wobblers were systematically studied and the variation
trends of their wobbling frequencies were
confirmed~\cite{Q.B.Chen2014PRC}.

With the successes of the collective Hamiltonian, it is interesting
to extend its applications. In $^{135}$Pr~\cite{Matta2015PRL}, not
only the transverse wobbling mode, but also its transition to the
longitudinal wobbling were observed. The experimental observations
have already been investigated by tilted axis cranking (TAC) with
the Strutinsky micro-macro method and the PRM in
Ref.~\cite{Matta2015PRL}. Here the collective Hamiltonian will be
applied to investigate the wobbling motions in $^{135}$Pr.

%%%%%%%%%%%%%%%%%%%%%%%%%%%%%%%%%%%%%%%%%%%%%%%%%%%%%%%%%%
%                    begin  framework
%%%%%%%%%%%%%%%%%%%%%%%%%%%%%%%%%%%%%%%%%%%%%%%%%%%%%%%%%%

\section{Theoretical framework}\label{sec2}

The adopted collective Hamiltonian was introduced in detail in
Refs.~\cite{Q.B.Chen2013PRC, Q.B.Chen2014PRC}. Choosing the azimuth
angle $\varphi$ as the collective coordinate, the collective
Hamiltonian reads
\begin{align}\label{eq4}
  \hat{H}_{\rm coll}=-\displaystyle \frac{\hbar^2}{2\sqrt{B(\varphi)}}
  \frac{\partial}{\partial\varphi}\frac{1}{\sqrt{B(\varphi)}}
  \frac{\partial}{\partial \varphi}+V(\varphi),
\end{align}
where the collective potential $V(\varphi)$ is extracted by
minimizing the total Routhian $E^\prime(\theta,\varphi)$ of TAC
calculations with respect to the polar angle $\theta$ for given
$\varphi$~\cite{Q.B.Chen2013PRC, Q.B.Chen2014PRC}. For a high-$j$,
the TAC Hamiltonian reads~\cite{Frauendorf1997NPA}
\begin{align}
  \hat{h}^\prime &=\hat{h}_{\rm def}-\bm{\omega}\cdot\hat{\bm{j}},\notag\\
  \bm{\omega}    &=(\omega\sin\theta\cos\varphi, \omega\sin\theta\sin\varphi,
                    \omega\cos\theta),
\end{align}
where $\hat{\bm{j}}$ is the single particle angular momentum and
$\hat{h}_{\rm def}$ is the single-$j$ shell Hamiltonian,
\begin{align}\label{eq7}
  \hat{h}_{\rm def}=\frac{1}{2}C\Big\{(\hat{j}_3^2-\frac{j(j+1)}{3})\cos\gamma
                   +\frac{1}{2\sqrt{3}}(\hat{j}_+^2+\hat{j}_-^2)\sin\gamma\Big\}.
\end{align}
In Eq.~(\ref{eq7}), the parameter $C$ is proportional to the
quadrupole deformation parameter $\beta$, and $\gamma$ is triaxial
deformation parameter. Diagonalizing the cranking Hamiltonian, one
ends up with the total Routhian
\begin{align}\label{eq9}
  E^\prime(\theta,\varphi)=\langle h^\prime\rangle-
  \frac{1}{2}\sum_{k=1}^3\mathcal{J}_k \omega_k^2, \quad
  \mathcal{J}_k: \textrm{moments of inertia},
\end{align}
and then the collective potential $V(\varphi)$.

To obtain the mass parameter, one can expand the collective
potential $V(\varphi)$ with respect to $\varphi$ at
$\varphi=0^\circ$ up to $\sim \varphi^2$ terms to extract the
stiffness parameter (labeled $K$) of
$V(\varphi)$~\cite{Q.B.Chen2014PRC}, and then
\begin{align}\label{eq6}
 B=\frac{K}{\Omega^2}
\end{align}
with $\Omega$ the wobbling frequency. For example, for a simple
wobbler, its stiffness parameter is
$K=\omega^2(\mathcal{J}_1-\mathcal{J}_2)$~\cite{Q.B.Chen2014PRC},
and its wobbling frequency can be calculated by the triaxial rotor
model:~\cite{Bohr1975}
\begin{align}
 \hbar\Omega=\hbar\omega\sqrt{\frac{(\mathcal{J}_1-\mathcal{J}_2)
 (\mathcal{J}_1-\mathcal{J}_3)}{\mathcal{J}_3\mathcal{J}_2}},
\end{align}
with $\omega$ the rotational frequency. Thus, according to
Eq.~(\ref{eq6}), the mass parameter is~\cite{Q.B.Chen2014PRC}
\begin{align}\label{eq5}
 B=\frac{\mathcal{J}_2\mathcal{J}_3}{\mathcal{J}_1-\mathcal{J}_3}.
\end{align}

\begin{figure}[!h*]
  \begin{center}
    \includegraphics[width=9 cm]{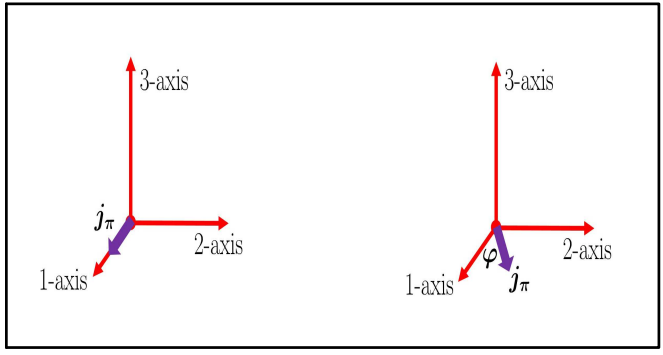}
    \caption{(Color online) Sketch of the angular momentum vector of the proton
    particle with respect to the principal axis frame.}\label{fig1}
  \end{center}
\end{figure}

For an odd-$A$ wobbler, one further introduces the harmonic frozen
alignment (HFA) approximation~\cite{Frauendorf2014PRC,
Frauendorf2015PRC}; i.e., the odd particle is assumed to be firmly
aligned with axis 1 (see left panel of Fig.~\ref{fig1}), and its
angular momentum is considered as a constant number $j$. Such an
assumption leads to an $\omega$-dependent effective MOI for axis 1
with $j/\omega$. Therefore, the Eq.~(\ref{eq5}) is replaced
by~\cite{Q.B.Chen2014PRC}
\begin{align}\label{eq1}
 B(\omega)=\frac{\mathcal{J}_2\mathcal{J}_3}{\mathcal{J}_1^*(\omega)-\mathcal{J}_3},
 \quad \mathcal{J}_1^*(\omega)=\mathcal{J}_1+\frac{j}{\omega}.
\end{align}

If the angular momentum of the odd particle tilts from axis 1 toward
axis 2, as illustrated in the right panel of Fig.~\ref{fig1}, the
effective MOI induced should be modified accordingly. If the tilted
angle is $\varphi$, the effective MOIs for axes 1 and 2 are
\begin{align}
\label{eq3}
 \mathcal{J}_1^*(\omega)&=\mathcal{J}_1+\frac{j\cos\varphi}{\omega},\\
 \mathcal{J}_2^*(\omega)&=\mathcal{J}_2+\frac{j\sin\varphi}{\omega}.
\end{align}
Correspondingly, the mass parameter (\ref{eq1}) should be rewritten as
\begin{align}\label{eq2}
 B(\omega)=\frac{\mathcal{J}_2^*(\omega)
 \mathcal{J}_3}{\mathcal{J}_1^*(\omega)-\mathcal{J}_3}.
\end{align}

With the collective potential from the TAC
model~\cite{Q.B.Chen2013PRC, Q.B.Chen2014PRC} and the mass parameter
from the HFA formula~(\ref{eq2}), the collective Hamiltonian
(\ref{eq4}) is constructed. Similar to Refs.~\cite{Q.B.Chen2013PRC,
Q.B.Chen2014PRC}, the collective Hamiltonian is solved by
diagonalization. Since the collective Hamiltonian is invariant with
respect to the $\varphi \to -\varphi$ transformation, one chooses
the following bases
\begin{align}
  \psi_{n}^{(1)}(\varphi)
  &=\sqrt{\frac{2}{\pi(1+\delta_{n0})}}
  \frac{\cos 2n\varphi}{B^{1/4}(\omega)}, \quad n\geq 0,\\
  \psi_{n}^{(2)}(\varphi)
  &=\sqrt{\frac{2}{\pi}}
  \frac{\sin 2n\varphi}{B^{1/4}(\omega)}, \quad n\geq 1,
\end{align}
which satisfy
\begin{align}
 \psi_{n}^{(1)}(-\varphi)=\psi_{n}^{(1)}(\varphi), \quad
 \psi_{n}^{(2)}(-\varphi)=-\psi_{n}^{(2)}(\varphi),
\end{align}
and the periodic boundary condition as
\begin{align}
 \psi_{n}^{(1)}(\varphi)=\psi_{n}^{(1)}(\varphi+\pi), \quad
 \psi_{n}^{(2)}(\varphi)=\psi_{n}^{(2)}(\varphi+\pi).
\end{align}

\section{Numerical details}\label{sec3}

In the following calculations, the configuration of the wobbling
bands in $^{135}$Pr is adopted as $\pi (1h_{11/2})^1$. The
quadrupole deformation parameters follow
Refs.~\cite{Frauendorf2014PRC, Matta2015PRL} as $\beta=0.17$ and
$\gamma=-26.0^\circ$. Accordingly, the axes 1, 2, and 3 are
respectively the short, intermediate, and long axes. The MOIs for
the three principal axes are taken as $\mathcal{J}_1$,
$\mathcal{J}_2$, $\mathcal{J}_3$=13.0, 21.0,
4.0~$\hbar^2/\rm{MeV}$~\cite{Frauendorf2014PRC}. It is seen that all
the parameters are the same as in previous
works~\cite{Frauendorf2014PRC, Matta2015PRL}, and no adjustable
parameters are introduced in the present calculations.

%%%%%%%%%%%%%%%%%%%%%%%%%%%%%%%%%%%%%%%%%%%%%%%%%%%%%%%%%%
%                    begin  results and discussion
%%%%%%%%%%%%%%%%%%%%%%%%%%%%%%%%%%%%%%%%%%%%%%%%%%%%%%%%%%

\section{Results and discussion}\label{sec4}

In a recent reported transverse wobbling partners in the $A \sim
130$ mass region, $^{135}$Pr, the wobbling frequency decreases with
spin, and the $\Delta I=1$ interband transitions between the partner
bands display primarily $E2$ character~\cite{Matta2015PRL}. In
Refs.~\cite{Matta2015PRL, Frauendorf2014PRC}, the TAC Strutinsky
micro-macro calculations adopt the deformation parameters
$\beta=0.17$ and $\gamma=-26.0^\circ$ and the PRM (or so-called
quasiparticle triaxial rotor model) adopts the MOIs as
$\mathcal{J}_1$, $\mathcal{J}_2$, $\mathcal{J}_3$=13.0, 21.0,
4.0~$\hbar^2/\rm{MeV}$, respectively. In the present collective
Hamiltonian calculations, we also use the same
parameters~\cite{Matta2015PRL, Frauendorf2014PRC}, and no additional
parameters.

\begin{figure}[!th]
  \begin{center}
    \includegraphics[width=4 cm]{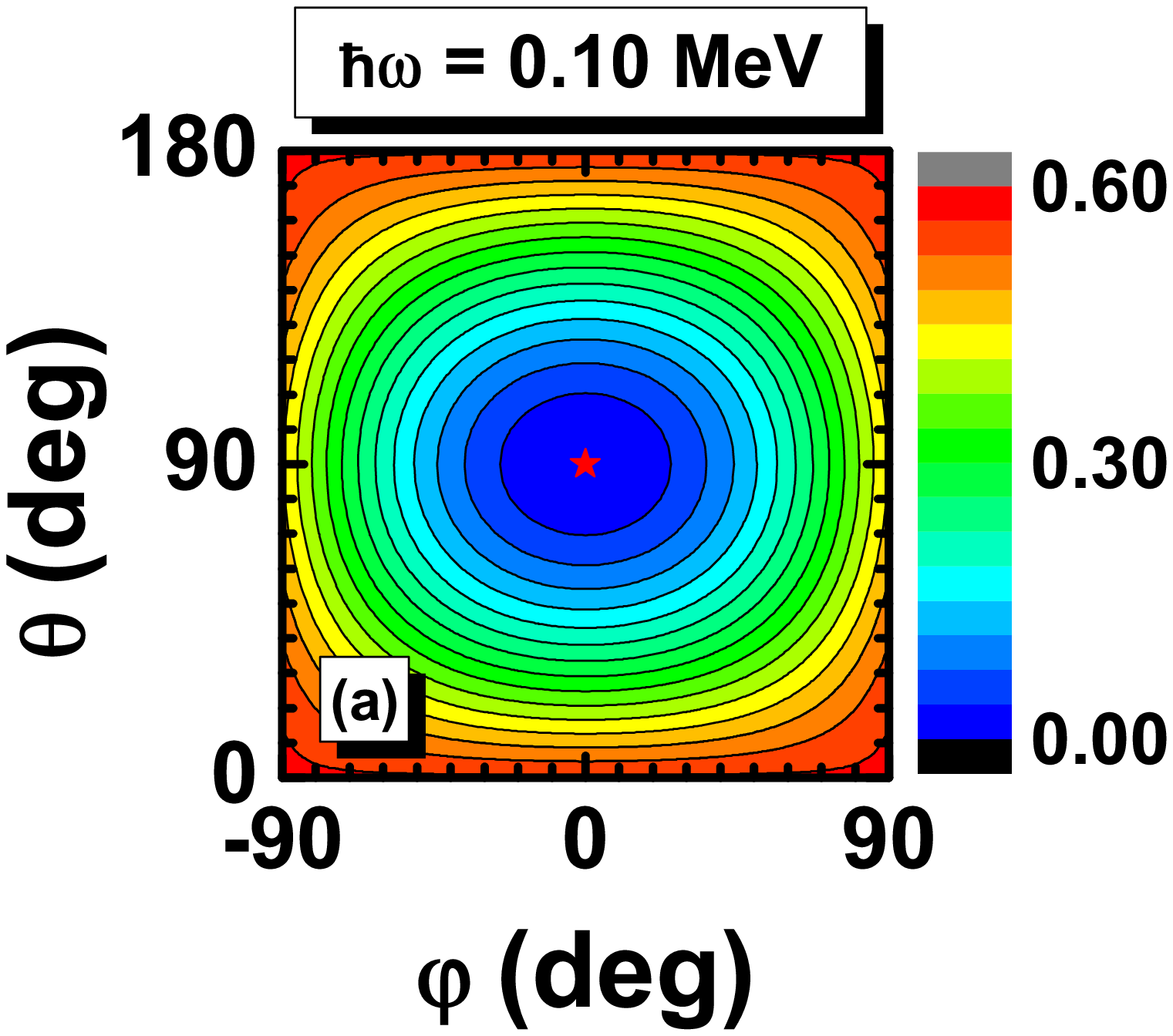}
    \includegraphics[width=4 cm]{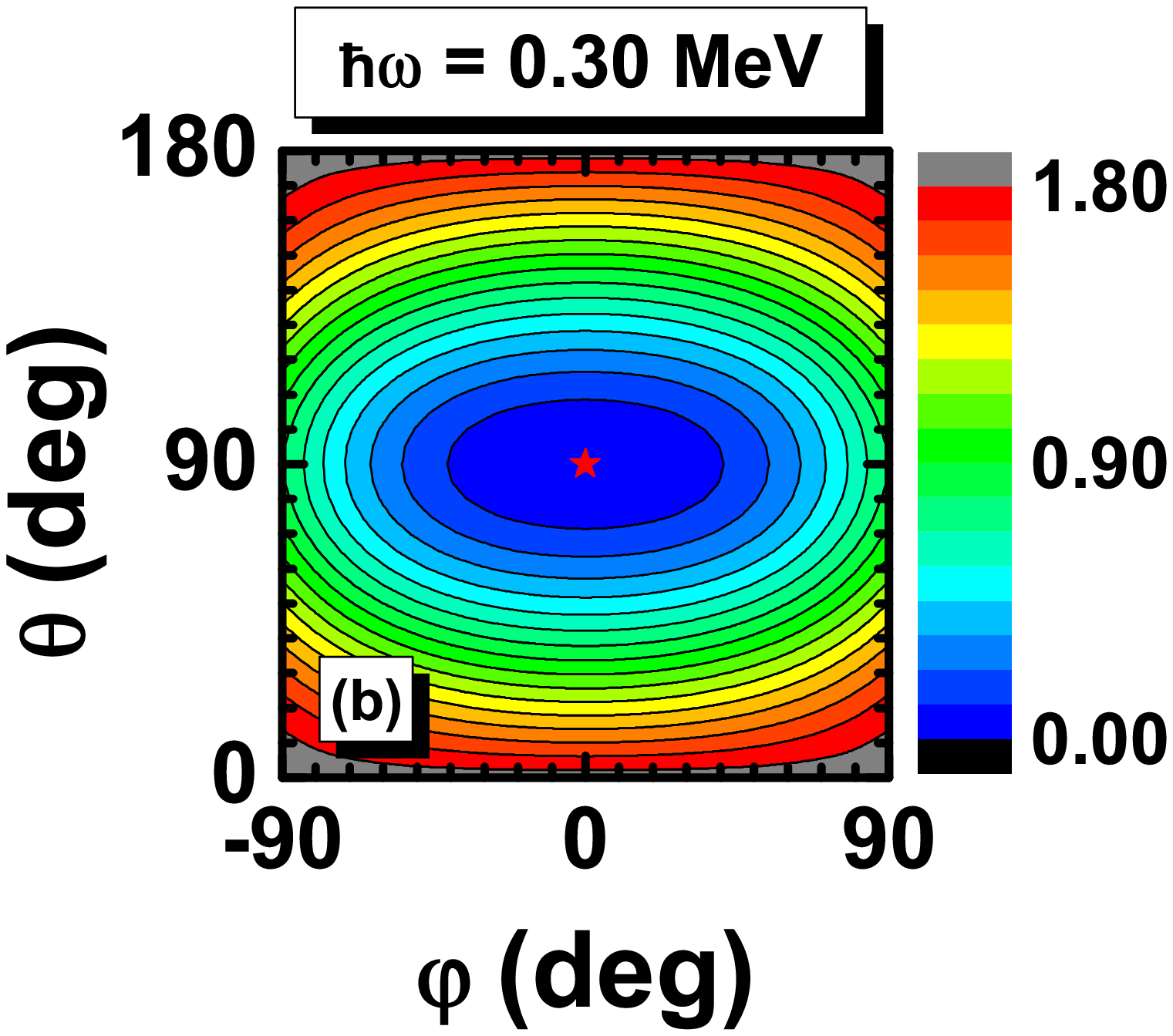}
    \includegraphics[width=4 cm]{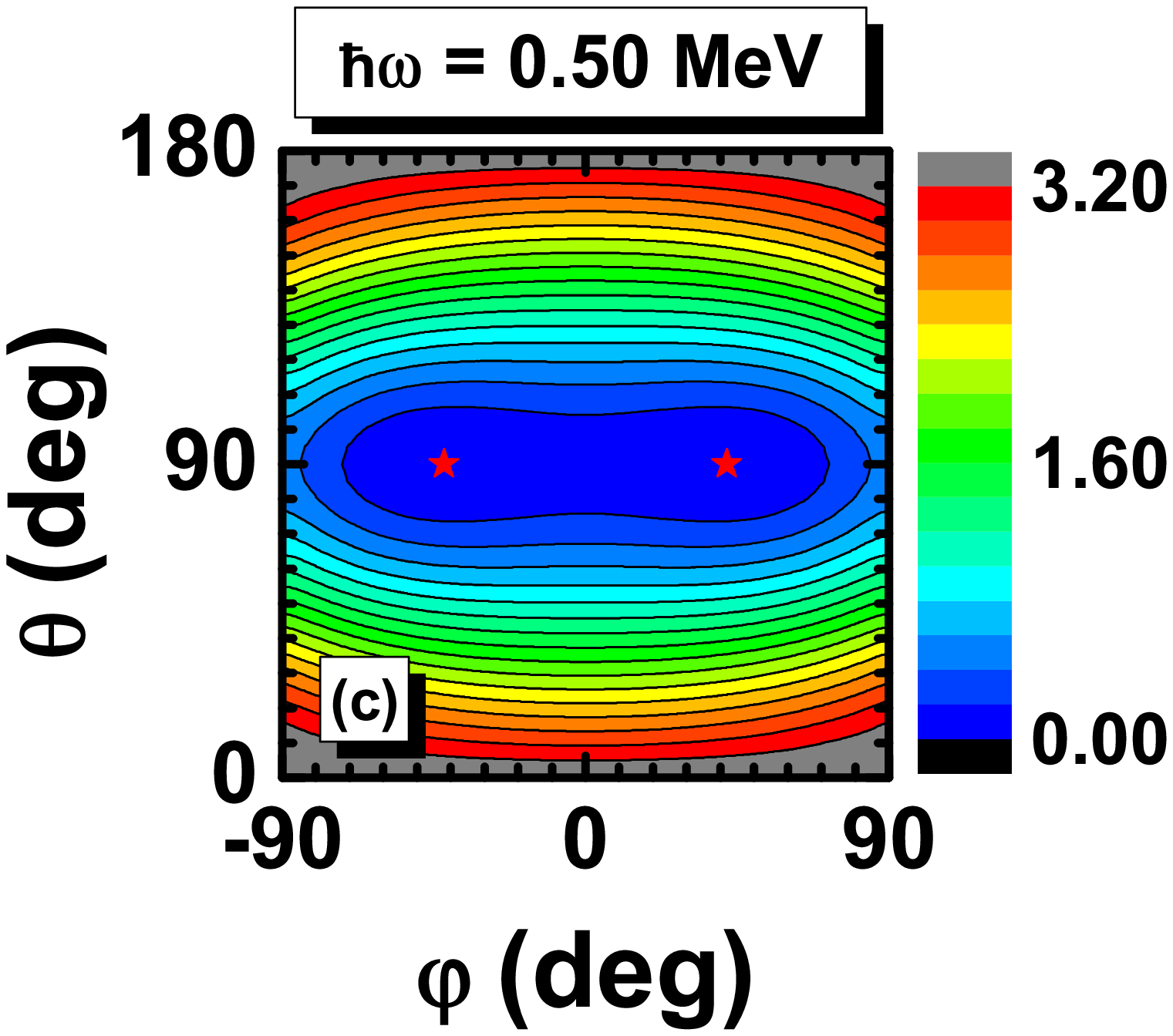}
    \includegraphics[width=4 cm]{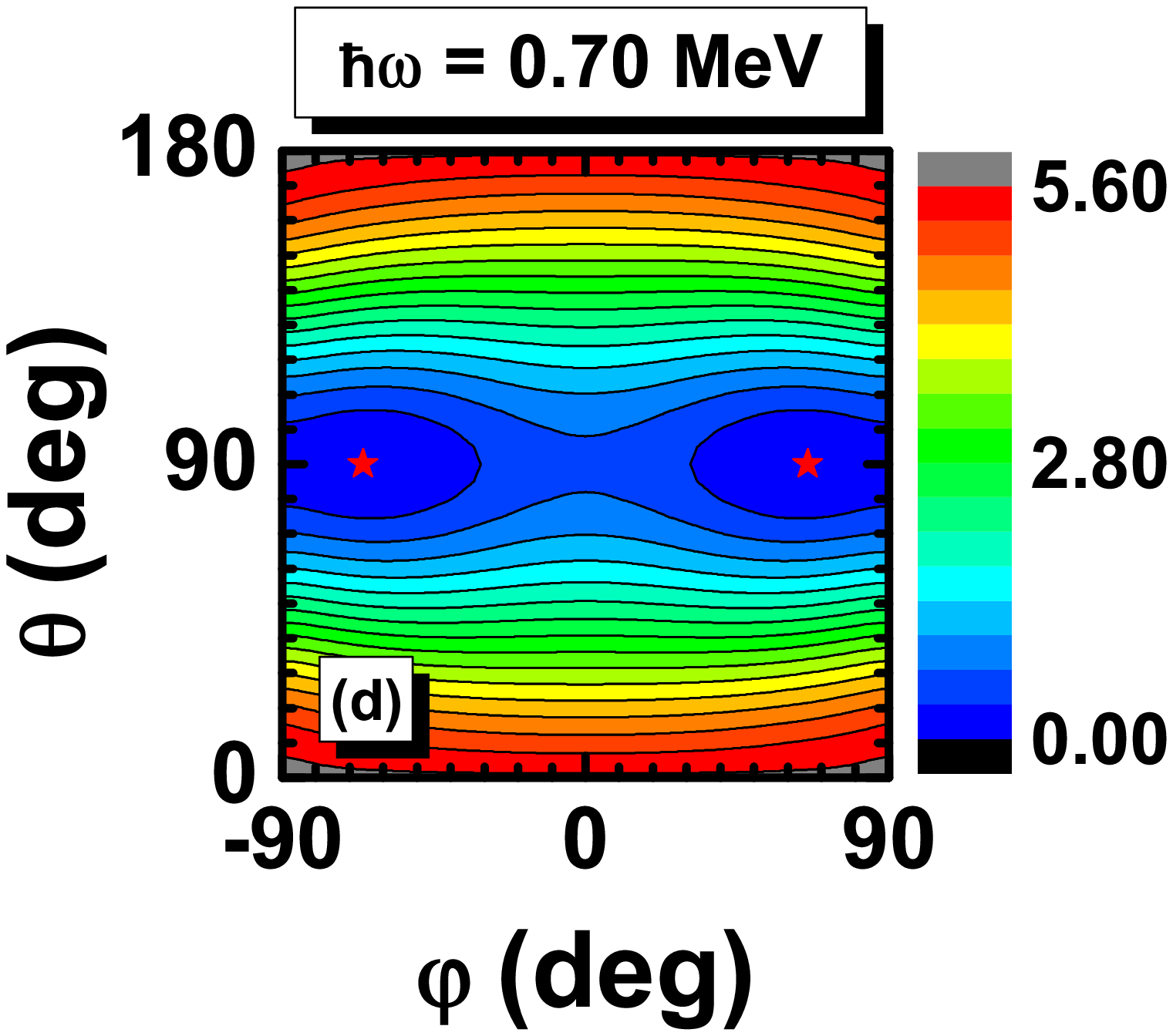}
    \caption{(Color online) Contour plots of the total Routhian surface
    calculation $E^\prime(\theta,\varphi)$ for $^{135}$Pr at the frequencies
    $\hbar\omega=0.10$, $0.30$, $0.50$, and $0.70~\rm MeV$. All energies at each
    rotational frequency are normalized with respect to the absolute
    minimum.}\label{fig2}
  \end{center}
\end{figure}

The total Routhian surfaces $E^\prime(\theta,\varphi)$ calculated by
the TAC model for $^{135}$Pr at the rotational frequencies
$\hbar\omega=0.10$, $0.30$, $0.50$, and $0.70~\rm MeV$ are shown in
Fig.~\ref{fig2}, where the minima are labeled with red stars. It can
be seen that all the total Routhian surfaces are symmetric with
respect to the $\varphi =0^\circ$ and $\theta=90^\circ$ lines, as a
result of the invariance of the intrinsic quadrupole moments with
respect to the $D_2$ symmetry.

It is shown that the $\theta$ values of the minima always locate at
$\theta=90^\circ$. This is because axis 3 is of the smallest MOI
and, as a consequence, the angular momentum prefers to align in the
1-2 plane. With the increase of rotational frequency, the $\varphi$
values of the minima gradually deviate from a vanishing value to
finite angles. As a result, the number of the minima changes from
one to two. This implies the rotational mode changes from a
principal axis rotation at low frequencies (e.g., $\hbar\omega=0.10$
and $0.30$ MeV) to a planar rotation at high frequencies (e.g.,
$\hbar\omega=0.70$ and $0.90$ MeV). These features provide a hint of
the existence of the transverse wobbling
mode~\cite{Q.B.Chen2014PRC}.

\begin{figure}[!ht]
  \begin{center}
    \includegraphics[width=10 cm]{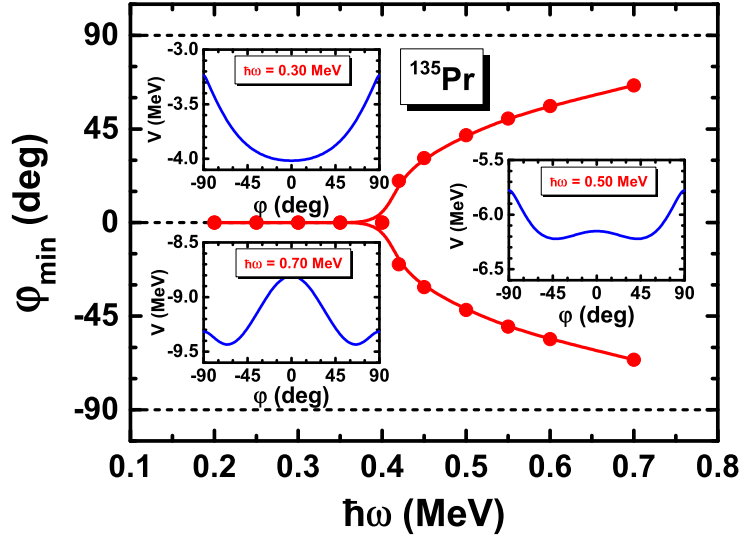}
    \caption{(Color online) $\varphi_{\min}$, i.e., the $\varphi$ which
    minimizes the total Routhian surface, as a function of rotational frequency and
    the extracted collective potential $V(\varphi)$ at $\hbar\omega=0.30$, $0.50$,
    and $0.70~\rm{MeV}$.}\label{fig3}
  \end{center}
\end{figure}

To see more clearly, $\varphi_{\min}$, i.e., the $\varphi$ which
minimizes the total Routhian surface, is shown in Fig.~\ref{fig3} as
a function of rotational frequency. $\varphi_{\min}$ is zero below
$\hbar\omega = 0.40~\mathrm{MeV}$, and is bifurcate above this
rotational frequency. Thus $\hbar\omega = 0.40~\mathrm{MeV}$ is the
critical rotational frequency at which the rotational mode changes.
For $\hbar\omega > 0.40~\mathrm{MeV}$, $\varphi_{\min}$ gradually
deviates from zero and, at $\hbar\omega=0.70~\rm{MeV}$, reaches
$\sim \pm 65^\circ$. It is expected that it would approach to $\pm
90^\circ$ with the increasing rotational frequency. In that case,
the rotational mode changes from a planar rotation to a principal
axis rotation around axis 2.

In Fig.~\ref{fig3}, we also show the collective potential
$V(\varphi)$ obtained by minimizing the total Routhian
$E^{\prime}(\theta,\varphi)$ with respect to $\theta$ for a given
$\varphi$ at $\hbar\omega=0.30$, $0.50$, and $0.70~\rm{MeV}$. As the
rotational frequency increases, $V(\varphi)$ changes from potential
shaped like a harmonic oscillator, with one minimum at
$\varphi_{\min}=0^\circ$ for $\hbar\omega=0.30$ MeV, to one shaped
like a sombrero, with two identical minima at $\varphi_{\min}\neq
0^\circ$ for $\hbar\omega=0.50$ and 0.70 MeV. The two symmetric
minima are separated by a potential barrier. The height of the
barrier can be defined as $\Delta V=V(0)-V(\varphi_{\min})$. It is
found that $\Delta V$ increases with rotational frequency, e.g.,
from 0.09 MeV at $\hbar\omega=0.50$ MeV to 0.65 MeV at
$\hbar\omega=0.70$ MeV. It is expected that, if the rotational
frequency continuously increases, $\Delta V$ would become larger and
drive the minima to approach $\pm 90^\circ$, which then changes the
rotational mode from a planar rotation to a principal axis rotation
around axis 2.

\begin{figure}[!ht]
  \begin{center}
    \includegraphics[width=8 cm]{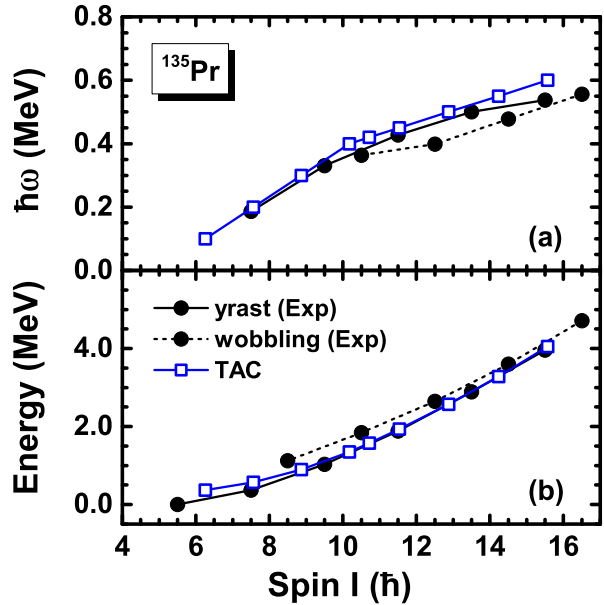}
    \caption{(Color online) Rotational frequency (upper panel) and
    rotational energy spectra (lower panel) of the yrast band in $^{135}\rm{Pr}$
    as functions of the angular momentum calculated
    by TAC (open squares) in comparison with the data (solid dots)
    of Ref.~\cite{Matta2015PRL}. In the TAC calculations, the quantal
    correction $1/2$ has been extracted for the angular
    momenta~\cite{Frauendorf1996Z.Phys.A}.}\label{fig4}
  \end{center}
\end{figure}

The obtained energy spectra and the $\hbar\omega$-$I$ relation from
the TAC are given in Fig.~\ref{fig4}, in comparison with the
experimental values of yrast band as well as the wobbling
band~\cite{Matta2015PRL}. In TAC, the spin $I$ is calculated with
the quantal correction $1/2$,
$I=J-1/2$~\cite{Frauendorf1996Z.Phys.A}, where $J$ is
$J=\sqrt{J_1^2+J_2^2+J_3^2}$ with $J_k$ the sum of the angular
momenta of the particle $j_{\pi k}=\langle \hat{j}_{\pi k}\rangle$
and the rotor $R_k=\mathcal{J}_k\omega_k$ as $J_k=j_{\pi k}+R_k$.
The energy spectra are calculated by $E=E^\prime+\omega J$. It is
shown that both the $\hbar\omega$-$I$ relation and energy spectra of
the yrast band are well reproduced by the TAC calculations. There is
a kink in the $I$-$\hbar\omega$ relation at $\hbar\omega=0.4$ MeV
($\sim 10\hbar$). This is attributed to the reorientation of the
core angular momentum from axis 1 toward axis 2, as shown in
Fig.~\ref{fig3}. As the wobbling band cannot be given by the TAC
calculations, the collective Hamiltonian method will be applied.

\begin{figure}[!th]
  \begin{center}
    \includegraphics[width=8 cm]{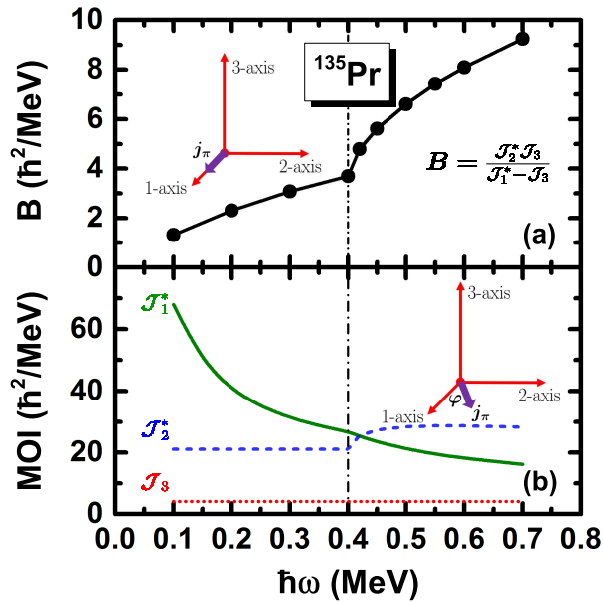}
    \caption{(Color online) The calculated mass parameter (upper panel)
    as well as the effective MOIs of the three principal axes
    $\mathcal{J}_1^*$, $\mathcal{J}_2^*$, and $\mathcal{J}_3$ (lower panel)
    as a function of rotational frequency $\hbar\omega$.
    A diagrammatic sketch of the angular momentum vector of the proton
    particle with respect to the principal axis frame is also shown.}\label{fig5}
  \end{center}
\end{figure}

The mass parameter in the collective Hamiltonian is calculated by
the HFA approximation formula~(\ref{eq2}), where the effective MOIs
induced by the proton particle are taken into account. The obtained
mass parameter as well as the effective MOIs of the three principal
axes are shown in Fig.~\ref{fig5} as functions of rotational
frequency. It is seen that the MOI of axis 3, $\mathcal{J}_3$,
remains constant, as the proton particle angular momentum has no
component along the axis 3 in the HFA approximation. The effective
MOI of axis 2, $\mathcal{J}_2^*$, is a constant at $\hbar\omega\leq
0.40~\rm{MeV}$, and increases after $\hbar\omega = 0.40~\rm{MeV}$.
The reason is that the proton particle angular momentum deviates
from axis 1, moving toward axis 2 at $\hbar\omega>0.40$ MeV. The
effective MOI of axis 1, $\mathcal{J}_1^*$, decreases with
rotational frequency due to the factor $1/\omega$ in
Eq.~(\ref{eq3}). As a consequence, the mass parameter increases with
the rotational frequency as shown in Fig.~\ref{fig5}(a). At
$\hbar\omega=0.40$ MeV there is a kink, corresponding to the
transition from principal axis rotation to planar rotation.

\begin{figure}[!ht]
  \begin{center}
    \includegraphics[width=7 cm]{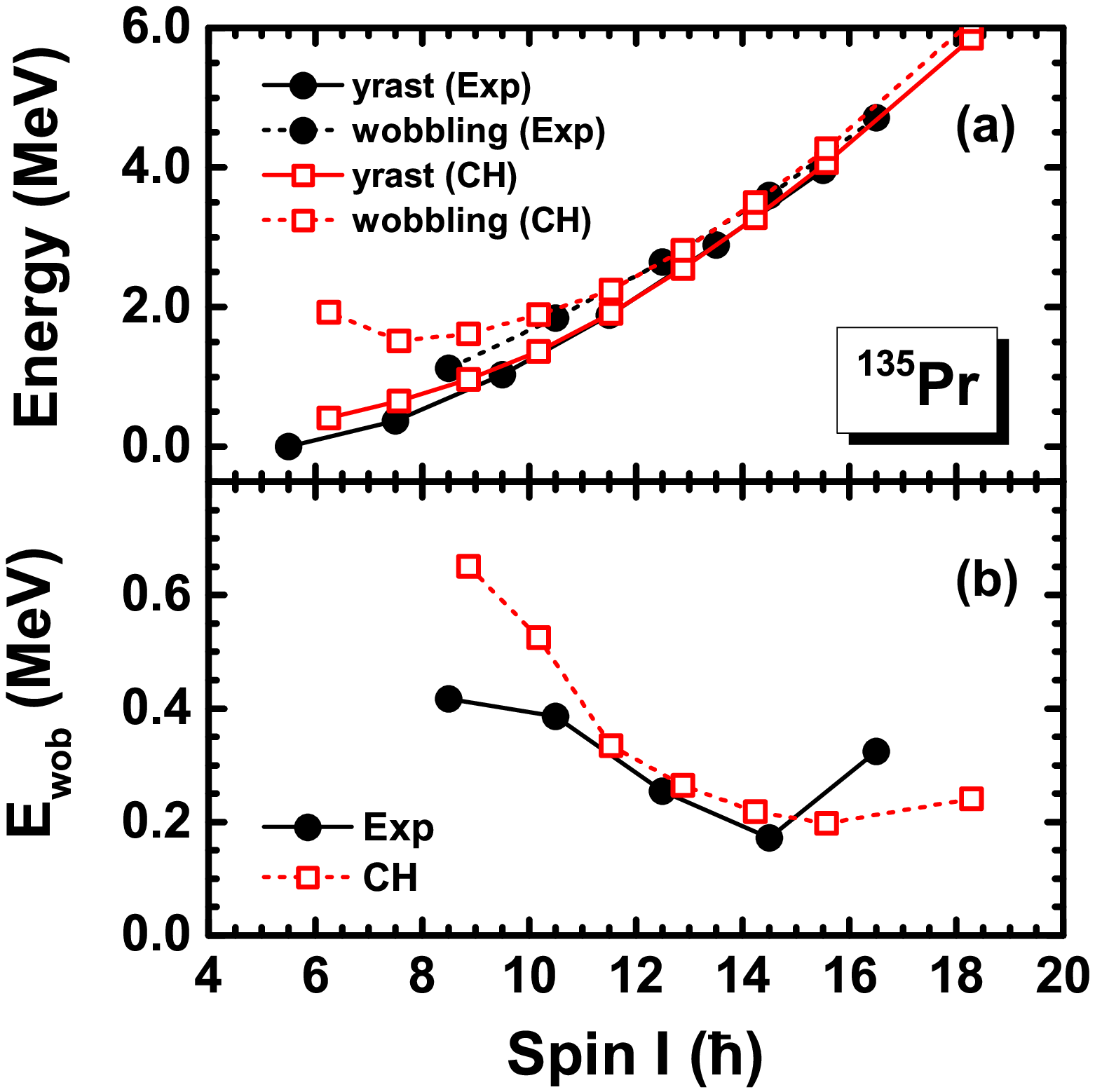}~~
    \includegraphics[width=7 cm]{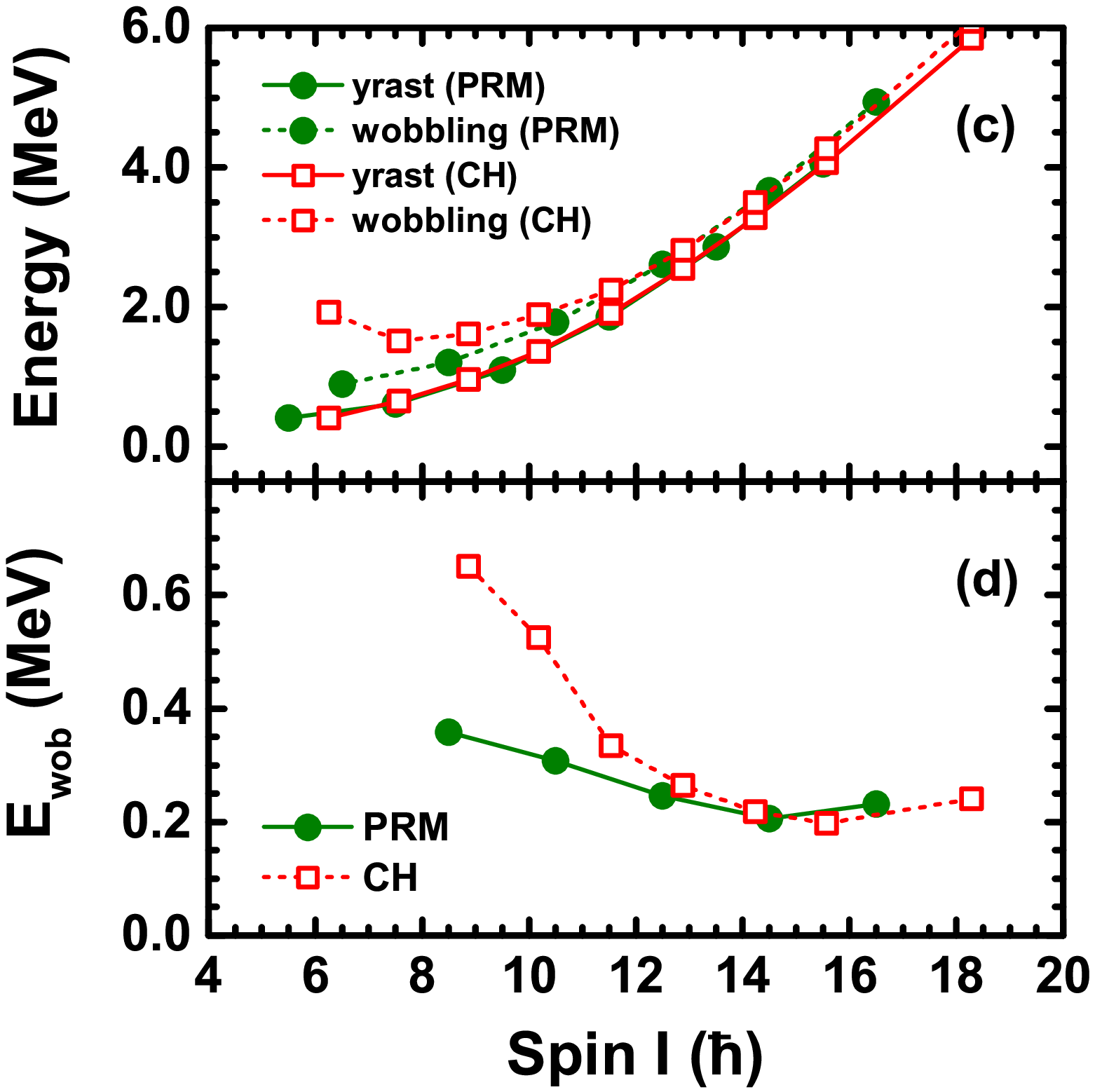}
    \caption{(Color online) Energy spectra of the yrast and wobbling
    bands (a) and the corresponding wobbling frequency
    (b) in $^{135}\rm{Pr}$ as functions of the angular
    momentum calculated by the collective Hamiltonian in comparison with the
    data of Ref.~\cite{Matta2015PRL}. In the collective
    Hamiltonian results, the angular momenta are calculated
    from the TAC model. Similar comparisons with PRM are shown
    in (c) and (d).}\label{fig6}
  \end{center}
\end{figure}

After obtaining the collective potential and the mass parameter, the
collective Hamiltonian (\ref{eq4}) is constructed. The
diagonalization of the collective Hamiltonian yields the collective
energy levels and the corresponding collective wave functions. The
lowest collective level at each cranking frequency corresponds to
the yrast mode, and the second lowest one corresponds to the
one-phonon wobbling excitation~\cite{Q.B.Chen2014PRC}. They are
compared with the data~\cite{Matta2015PRL} in Fig.~\ref{fig6}(a),
and good agreement can be seen.

From the energy spectra, the wobbling frequency $E_{\rm wob}$ is
extracted by calculating the energy difference between the yrast and
wobbling bands. The obtained $E_{\rm wob}$ as a function of spin is
shown in Fig.~\ref{fig6}(b), in comparison with the
data~\cite{Matta2015PRL}. At $I\leq 14.5\hbar$, both the theoretical
and experimental wobbling frequencies decrease with spin, which
provides the evidence of transverse wobbling motion. The theoretical
calculations overestimate the data at $I<10.5\hbar$. The reason
might be attributed to the fact that the HFA approximation used to
derive the mass parameter is not a good approximation at low
spins~\cite{Q.B.Chen2014PRC}. At the high spin region ($I\geq
14.5\hbar$), the experimental wobbling frequency shows an increasing
trend, indicating the wobbling mode transition from transverse to
longitudinal type~\cite{Matta2015PRL}. The collective Hamiltonian
calculations well reproduce this transition.

As mentioned in the Introduction, the PRM solutions for $^{135}$Pr
have been given in Refs.~\cite{Frauendorf2014PRC, Matta2015PRL}. In
Figs.~\ref{fig6}(c) and~\ref{fig6}(d), the energy spectra and
wobbling frequency obtained by the collective Hamiltonian are
compared with those by the PRM. It is seen that the collective
Hamiltonian can well reproduce the PRM energy spectra, except the
first two states in the wobbling band. Also, for the wobbling
frequency, the collective Hamiltonian has good agreement with the
PRM in the high spin region ($I\geq 12.5\hbar$), but overestimates
in the low spin region ($I \leq 10.5\hbar$). This implies that the
approximation used in the present collective Hamiltonian in the high
spin region works better than that in the low spin region.

The transition of the wobbling mode can be understood from the
effective MOIs, $\mathcal{J}_k^*$. As shown in Fig.~\ref{fig5}(b),
$\mathcal{J}_1^*$ is much larger than $\mathcal{J}_2^*$ and
$\mathcal{J}_3$ at $\hbar\omega \leq 0.40~\rm{MeV}$. As a result,
the total angular momentum favors axis 1. This corresponds to
rotation about the short axis (axis 1) and forms the transverse
wobbling mode. In the large rotational frequency region, however,
$\mathcal{J}_2^*$ becomes larger than $\mathcal{J}_1^*$ and
$\mathcal{J}_3$. This leads to the tilt of the total angular
momentum toward axis 2, and the transverse wobbling mode changes to
the longitudinal wobbling mode.

\begin{figure}[!ht]
  \begin{center}
    \includegraphics[width=11 cm]{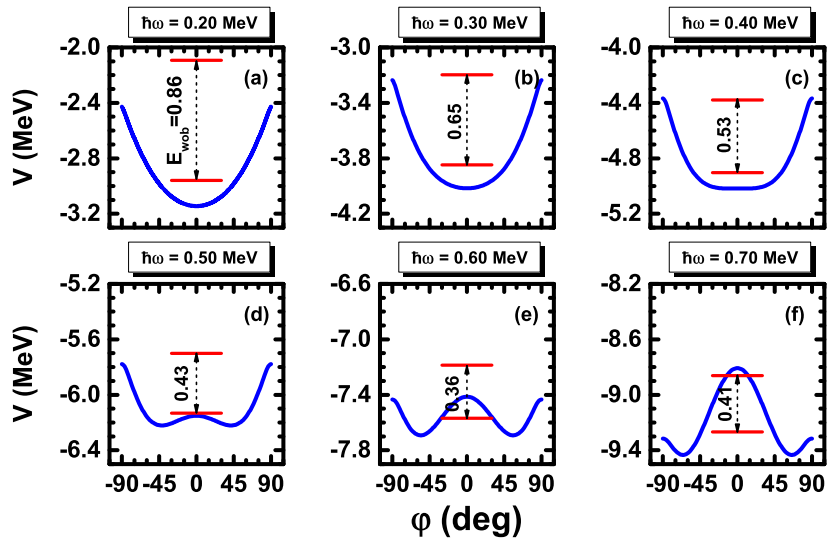}
    \caption{(Color online) Collective potential calculated by
    TAC model and two lowest collective energy levels obtained from
    the collective Hamiltonian at rotational frequencies
    $\hbar\omega=0.20$-$0.70~\rm{MeV}$. The wobbling frequency $E_{\rm wob}$
    for each rotational frequency is also shown.}\label{fig7}
  \end{center}
\end{figure}

It is interesting to understand the variation of the wobbling
frequency from the calculations of the collective Hamiltonian. In
Fig.~\ref{fig7}, the collective potentials as well as the obtained
yrast and wobbling energy levels at rotational frequencies
$\hbar\omega=0.20$, 0.30, 0.40, 0.50, 0.60, and $0.70~\rm{MeV}$ are
shown. The wobbling frequency $E_{\rm wob}$ for each rotational
frequency is also presented. For $\hbar\omega \leq 0.40~\rm{MeV}$,
the collective potential is of a harmonic oscillator shape with its
bottom part becoming flatter with the increase of rotational
frequency. This, in combination with the increase of the mass
parameter [see Fig.~\ref{fig5}(a)], makes the wobbling excitation
easier, and thus the wobbling frequency decreases, e.g., from
$E_{\rm wob}=0.86$ MeV at $\hbar\omega=0.20$ MeV to $E_{\rm
wob}=0.53$ MeV at $\hbar\omega=0.40$ MeV. At $\hbar\omega=0.50$ and
0.60 MeV, there appear two symmetric minima and a potential barrier
between them. The continuous decrease of wobbling frequency is
attributed to the appearance and increase of the barrier, which will
suppress the tunneling probability between the two
minima~\cite{Q.B.Chen2014PRC}. When $\hbar\omega \geq 0.70$ MeV, the
minima of the collective potential gradually approach $\varphi=\pm
90^\circ$. The potential barriers at $\pm 90^\circ$ become much
lower than that at $0^\circ$, and will eventually disappear at a
large enough rotational frequency. As a result, the potential at
$\pm 90^\circ$ becomes stiffer, and the wobbling excitations become
harder. Thus, the wobbling frequency here shows an increasing trend.

\begin{figure}[!ht]
  \begin{center}
    \includegraphics[width=11 cm]{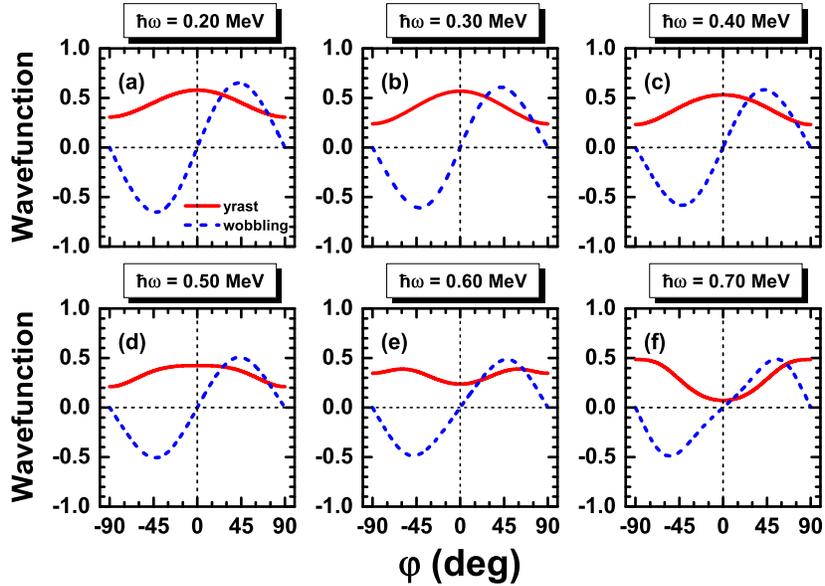}
    \caption{(Color online) Collective wave functions obtained from the collective
    Hamiltonian at rotational frequency $\hbar\omega=0.20$-$0.70~\rm{MeV}$.}\label{fig8}
  \end{center}
\end{figure}

The obtained wave functions of the yrast and wobbling bands at
different rotational frequencies are shown in Fig.~\ref{fig8}. It is
seen that the wave functions are symmetric for the yrast band and
antisymmetric for the wobbling band with respect to $\varphi \to
-\varphi$ transformation. Thus the broken signature symmetry in the
TAC model is restored in the collective Hamiltonian by the
quantization of wobbling angle $\varphi$ and the consideration of
quantum fluctuation along the $\varphi$ motion. The peak of the wave
function of yrast state is located at $\varphi=0^\circ$ at
$\hbar\omega\leq 0.40~\rm{MeV}$, and deviates from $\varphi=0^\circ$
at $\hbar\omega > 0.40~\rm{MeV}$. This reflects the transition from
the principal axis rotation to planar rotation. For the wave
functions of wobbling states, which correspond to one-phonon
excitations, they are odd functions and the values at
$\varphi=0^\circ$ and $\pm 90^\circ$ are all zero.

%%%%%%%%%%%%%%%%%%%%%%%%%%%%%%%%%%%%%%%%%%%%%%%%%%%%%%%%
%                    begin  summary
%%%%%%%%%%%%%%%%%%%%%%%%%%%%%%%%%%%%%%%%%%%%%%%%%%%%%%%%

\section{Summary and perspective}\label{sec5}

In summary, the collective Hamiltonian based on the TAC model is
applied to describe the recently observed wobbling bands in
$^{135}$Pr. The collective parameters in the collective Hamiltonian,
including the collective potential and the mass parameter, are
calculated by the TAC model and the HFA formula, respectively.

For the yrast band, the energy spectra together with the relations
between the spin and the rotational frequency can be reproduced by
the TAC model with the configuration $\pi(1h_{11/2})^1$. Beyond the
TAC mean field approximation, the collective Hamiltonian reproduces
the energy spectra of both the yrast and wobbling bands well. It is
confirmed that the wobbling mode in $^{135}$Pr changes from the
transverse to longitudinal one with the increase of rotational
frequency. This transition is understandable by analyzing the
effective MOIs of the three principal axes. It is pointed out that
the effective MOI caused by the valence particle is of importance
for forming different type of wobbling mode, and the softness and
shapes of the collective potential determine the variation trends of
the wobbling frequency.

Here, the collective Hamiltonian is constructed based on a simple
single-$j$ shell model. The success of the collective Hamiltonian
here guarantees its application for more realistic TAC calculations,
e.g., the TAC covariant density functional
theory~\cite{J.Meng2013FP, J.Meng2016PS, J.Meng2016book}. After such
a TAC model is implemented, the collective potential and the mass
parameters in the collective Hamiltonian can be obtained in a fully
microscopic manner. Works along this direction are in progress.

\section*{Acknowledgements}

This work was partly supported by the Chinese Major State 973
Program No. 2013CB834400, the National Natural Science Foundation of
China (Grants No. 11335002, No. 11375015, No. 11461141002, and No.
11621131001), and the China Postdoctoral Science Foundation under
Grants No. 2015M580007 and No. 2016T90007.

%%%%%%%%%%%%%%%%%%%%%%%%%%%%%%%%%%%%%%%%%%%%%%%%%%%%%%%%
%                  begin refereee
%%%%%%%%%%%%%%%%%%%%%%%%%%%%%%%%%%%%%%%%%%%%%%%%%%%%%%%%

\end{CJK}

\end{document}